\documentclass[11pt]{article}     

\begin{document}           

\begin{center}

{\LARGE A Tiled-Table Convention for Compressing FITS Binary Tables}\\
\medskip
William Pence, NASA/GSFC\\
Rob Seaman, NOAO\\
Richard L. White, STScI\\
\medskip
Version 1.0\\
October 28, 2010\\

\end{center}

\section{Overview}

This document describes a convention for compressing FITS binary
tables that is modeled after the FITS tiled-image compression method
(White et al. 2009) that has been in use for about a decade. The input
table is first optionally subdivided into tiles, each containing an
equal number of rows,  then every column of data within each tile is
compressed and stored as a variable-length array of bytes in the
output FITS binary table.  All the header keywords from the input
table are copied to the header of the  output table and remain
uncompressed for efficient access. The output compressed  table
contains the same number and order of columns as in the input 
uncompressed binary table. There is one row in the output table
corresponding to each tile of rows in the input table.  In principle,
each column of data can be compressed using a different algorithm 
that is optimized for the type of data within that column, however in
the prototype implementation described here, the gzip algorithm is
used to compress every column.  

When compressing most FITS tables, a lossless algorithm will be
required in order to exactly preserve the values in each column. In
principle, however, a lossy compression algorithm could be used to
achieve higher compression in cases where the values in a particular
column do not need to be exactly preserved.

This convention currently can only be used to compress FITS binary
tables and is not applicable to FITS ASCII tables.

\section{Algorithm Details}

The algorithm for compressing a FITS binary table consists of 
the following sequence of steps: 

\begin{enumerate} 

\item[A. ]{\bf Divide Table into Tiles (Optional)}

In order to limit the amount of data that must be manipulated at one time,
large FITS tables may be optionally divided into tiles, where each tile
contains the same number of rows from the input table, except for the last
tile which may contain fewer rows. Each tile is compressed in turn, and is
stored in a row in the output compressed table. This convention places no
restriction on the tile size, but in practice, it is recommended that FITS
tables that are larger than  about 10 MB in size should be divided into
tiles so as to not impose too large of a memory resource burden on
software that must uncompress the table.  

The number of rows of data in each tile is defined by the {\tt ZTILELEN}
keyword, as described in Section 3.  If this keyword is not present, then
it should be assumed that the entire table has been compressed as a single
tile.

\item[B. ]{\bf Transpose the Rows and Columns}

The data cells in a FITS binary table are natively stored in row by row
order, where the values in the first row for each  column are given in
order, followed by the values for the second row, and so on. Because
binary tables can contain very heterogeneous types of data in different
columns, it can be difficult to directly compress this native stream of
data values.  One can almost always obtain better compression by
internally transposing the table values into `column-major' order, where
all the values for the first column occur first, followed by all the
values for the second column, and so on. Then each column of values can be
compressed separately, and, at least in principle, a different compression
algorithm may be chosen for each column that is optimized  for that
particular type of data. 

Note that fixed-length vector columns are transposed in the same way as
scalar columns.  For example, if the table contains a {\tt TFORMn = '30I'}
column (an array of 30 16-bit integers) and contains 50 rows, then the
transposed column will contain an array of 30 $\times$ 50 =  1500 16-bit
integers.

\item[C. ]{\bf Compress each Column}

Each column of data is compressed with a suitable compression algorithm. 
If the table has been divided into tiles, then the same compression
algorithm must be used for a given column in all the tiles. The mnemonic
name of the compression algorithm that is used for each column is given by
the corresponding {\tt ZCTYPn} keyword, as described in Section 3.

\item[D. ]{\bf Store the Compressed Bytes}

The compressed stream of bytes for each column is stored in the
corresponding  column in the output table.  The compressed table will have
exactly the same number and order of columns as in the input table, however
the data type of the columns in the output table will all have a
variable-length byte data type, with TFORMn = '1PB', which is appropriate
for storing the compressed byte stream. Each row in the output compressed
table corresponds to a tile of rows in the input uncompressed table. If the
input table is compressed as a single tile, then the output table will only
contain one row.    

\end{enumerate}

\section{Keywords}

With only a few exceptions, all the keywords from the uncompressed table
are copied verbatim, in order, into the header of the compressed table. 
The header keywords remain uncompressed in order to preserve fast read and
write access. Note in particular that the values of the  reserved column
descriptor keywords {\tt TTYPEn}, {\tt TUNITn}, {\tt TSCALn}, {\tt TZEROn},
{\tt TNULLn}, {\tt TDISPn}, and {\tt TDIMn}, as well as all the
column-specific WCS keywords defined in the FITS standard, have the same
values in both the original and in the compressed table,  with the
understanding that these keywords apply to the values in the {\em
uncompressed columns}. 

The only keywords that are not copied verbatim from the uncompressed table
header to the compressed table header are the mandatory {\tt NAXIS1}, {\tt
NAXIS2}, {\tt PCOUNT}, and {\tt TFORMn} keywords  (and possibly the
reserved  {\tt THEAP} keyword), which must necessarily describe the actual
structure of the compressed table.  The original values of these keywords
are stored in a new set of reserved keywords in the compressed table
header. The complete set of keywords that have a reserved meaning within
the header of a tile-compressed binary table are listed below:

\begin{itemize}

\item {\tt ZTABLE} (required keyword).  The value field of this keyword
shall contain the logical value T.  It indicates that the FITS binary
table extension contains a tile-compressed binary table.

\item {\tt ZTILELEN} (optional keyword).  The value of this  keyword shall
contain an integer representing the number of rows of data from the
original binary table that are contained in each tile of compressed data.
The number of rows in the last tile, which may be smaller than the previous
tiles, is given by {\tt ZNAXIS2 - ((NAXIS2 - 1) $\times$  ZTILELEN)}.  If
the {\tt ZTILELEN} keyword is absent, then it should be assumed that  the
whole table is compressed as a single tile;  the compressed table must
only have a single row in this case.

\item {\tt ZCTYPn}  (optional indexed keywords). The value field of these
keywords shall contain a character string giving the mnemonic name of the
algorithm that is used to compress and decompress column {\tt n} of the
table.  The current allowed values are {\tt GZIP\_1}, {\tt GZIP\_2}, and
{\tt RICE\_1}, and the corresponding algorithms are described in Section
4.  If this keyword is absent for a given column, then the default value
of {\tt GZIP\_2} should be assumed.

\item {\tt ZNAXIS1}  (required keyword). The value field of this keyword
shall contain an integer that gives the value of the {\tt NAXIS1} keyword
in the original FITS table header.  This represents the width in bytes of
each row in the uncompressed table.

\item {\tt ZNAXIS2} (required keyword).  The value field of this keyword
shall contain an integer that gives the value of the {\tt NAXIS2} keyword
in the original  FITS table header.  This represents the number of rows in
the uncompressed table.

\item {\tt ZPCOUNT} (required keyword). The value field of this keyword
shall contain an integer that gives the value of the {\tt PCOUNT} keyword
in the original FITS table header.

\item {\tt ZFORMn}  (required indexed keywords).  These required array
keywords  supply the character string value of the corresponding {\tt
TFORMn} keyword in the original FITS table header.

\item {\tt ZTHEAP} (optional keyword).  The value field of this keyword
shall contain an integer that gives the value of the {\tt THEAP} keyword,
if it exists, in the original FITS table header.  In practice, this
keyword is rarely used.

\item {\tt ZHEAPPTR} (optional keyword). The value field of this keyword
shall contain an integer that gives the byte offset from the start of the
main data table to the beginning of the compressed heap. Usage of this
keyword is described in Section 5.2.

\end{itemize} 

\section{Supported Compression Algorithms}

This section describes the currently defined compression algorithms.
Other compression algorithms may be added in the future.

\subsection{GZIP\_1 compression algorithm}

This compression algorithm is designated by the keyword {\tt ZCTYPn =
'GZIP\_1'}. Gzip is the compression algorithm used in the widely
distributed GNU free software utility of the same name.   It was created
by Jean-loup Gailly and Mark Adler.  It is based on the DEFLATE 
algorithm, which is a combination of LZ77 and Huffman coding. Further
information about this compression technique is readily available online
on the Web.  The ``gzip -1'' option is generally used which
significantly  improves the compression speed with only a small loss of
compression efficiency. 

\subsection{GZIP\_2 compression algorithm (default)}

This compression algorithm is designated by the keyword {\tt ZCTYPn =
'GZIP\_2'}. If the {\tt ZCTYPn} keyword is absent for a given column,
then this algorithm should be assumed. This algorithm is a variation of
the {\tt GZIP\_1} algorithm in which the bytes in the arrays of numeric
data columns are preprocessed by shuffling them so that they are
arranged in order of decreasing significance before being compressed.  
For example, a 5-element array of 2-byte (16-bit) integer values, with
an original byte order of
\begin{verbatim}
      A1 A2 B1 B2 C1 C2 D1 D2 E1 E2,
\end{verbatim}
will have the following byte order after shuffling the bytes:
\begin{verbatim}
      A1 B1 C1 D1 E1 A2 B2 C2 D2 E2.
\end{verbatim}
Byte shuffling is only performed for numeric binary table columns that
have {\tt TFORMn} data type codes of {\tt I, J, K, E, D, C,} or {\tt
M}.   The descriptors  for all variable-length array columns, with a {\tt
P} or {\tt Q} type code, are also byte-shuffled.  The bytes in columns
that  have a {\tt L, X,} or {\tt A} type code are not shuffled.

This byte-shuffling technique has been shown to be especially beneficial
when compressing floating-point values because the bytes containing the
exponent and the most significant bits of the  mantissa  are often  similar
for all the floating point values in the array.  Thus these repetitive byte
values generally compress very well when grouped together in this way. 
HDF Group has used this byte-shuffling technique when compressing
HDF5 data files (HDF 2000).

\subsection{RICE\_1 compression algorithm}

This compression algorithm is designated by the keyword {\tt ZCTYPn =
'RICE\_1'}. The Rice algorithm (Rice, R. F., Yeh, P.-S., and Miller, W. H.
1993,  in Proc. of the 9th AIAA Computing in Aerospace Conf.,
AIAA-93-4541-CP) is very simple and  fast.  It requires only enough memory
to hold a single block of 32 integers at a time and is able to adapt very 
quickly to changes  in the input array statistics.  This algorithm can only
be used to compress integer table columns.   Note that the byte-shuffling
technique that is used with the GZIP\_2 algorithm must not be applied to
the integer columns when using the Rice compression algorithm.. 

\section{Compressing Variable-Length Array Columns}

Compression of binary tables that contain variable-length array columns,
with a {\tt P} or {\tt Q} data type code, requires special consideration
because the table elements are not stored in the table directly, but
instead are stored in what is called the `data heap' which follows the main
table.  The columns in the main data table itself contains a `descriptor',
which is composed of 2 integers that give the size and location of the
array  in the heap.  Thus, in addition to compressing the descriptors
values, one must also compress the actual array values in the heap.

\subsection{Compression of the Descriptors}

The  descriptor values in the main table for {\tt P} or {\tt Q} type
variable-length array columns  consist of a pair of 32-bit integers or
64-bit integers, respectively.  For compression purposes, these columns 
are treated  the same as a  column with a {\tt TFORMn = '2J'} or {\tt
TFORMn = '2K'} data type.

\subsection{Heap compression}

After the main data table has been compressed,  the entire data heap in
the input table is compressed as a single block of data using the GZIP\_2
algorithm (described in Section 4.2).  The size of the data heap is given
by the {\tt PCOUNT} keyword (in the input table) and it begins with the
byte immediately following the main data table. (In fact, the {\tt THEAP}
keyword in the input table can be used to leave a gap between the end of
the main data table and the beginning of the actual heap, but this
technicality can be safely ignored here). To improve the compression
efficiency,  the bytes in each numeric variable-length array within the
heap are shuffled so that the most significant byte of each array element
is given first, followed by the next most significant byte of every
element, and so on.  This can be done efficiently by transversing every
variable-length array column in  the input table that has a {\tt I, J, K,
E, D, C,} or {\tt M} data type, in  order from the first row to the last.
(When uncompressing the heap, this byte unshuffling should be done in
reverse order to ensure that any overlapping arrays in the heap are
restored to their original byte order). The byte-shuffled heap is then
compressed with the gzip algorithm, and the resulting compressed stream
of bytes is stored in the heap of the output compressed table  {\em
immediately following} the previously written compressed bytes from  the
main data table.  The starting byte location of the compressed heap,
relative to the starting location of the main data table is given by the
reserved {\tt ZHEAPPTR} keyword. 

The value of the {\tt PCOUNT} keyword in the output table will give the
total size of the heap in the compressed table, which includes the size
of the compressed columns of data from the main data table plus the size
of the compressed heap.

\section{Prototype Implementation}

The advantages of tiled-table compression have been demonstrated 
on a small sample of astronomical FITS
binary tables, mostly from the HEASARC archive, using
a prototype version of the CFITSIO library that supports  this
compression method. (Tests on a wider
sample of tables are pending).  Each table was compressed in
three different ways.  In the first test, each FITS table was simply
compressed using the gzip utility program.   This is currently the most
commonly used method for compressing FITS tables.  For comparison, the
second test used the GZIP\_1 compression method where the rows and
columns in the table are transposed before compressing each column
using gzip.  Finally, the third test used the GZIP\_2 method where the
bytes  in all the numeric columns are shuffled into decreasing order of
significance, in addition to  transposing the rows and columns.  The
following table shows the compression ratio (original table size
divided by the compressed table size) for these 3 different compression
methods.  The 13 different FITS
tables have been sorted in order of increasing gzip compression ratio.

\begin{center}
\begin{tabular}{rrrr|c}
\hline
   File   & gzip & GZIP\_1 & GZIP\_2 &  Disk Savings factor \\
\hline
     1  &  1.11  & 1.16 &  1.49  &  3.3 \\
     2  &  1.41  & 1.72 &  2.02  &  1.7 \\
     3  &  1.48  & 1.70 &  1.94  &  1.5 \\
     4  &  1.51  & 1.81 &  2.12  &  1.6 \\
     5  &  1.61  & 1.95 &  2.34  &  1.5 \\
     6  &  1.71  & 1.89 &  1.95  &  1.2 \\
     7  &  1.73  & 2.43 &  2.57  &  1.4 \\
     8  &  1.83  & 2.14 &  3.08  &  1.5 \\
     9  &  2.63  & 2.63 &  3.75  &  1.2 \\
    10  &  2.88  & 3.93 &  6.35  &  1.3 \\
    11  &  3.07  & 4.88 &  3.13  &  1.0 \\
    12  &  3.85  & 5.51 & 22.59  &  1.3 \\
    13  &  5.39  & 7.63 &  7.60  &  1.1 \\
\hline
\end{tabular}
\end{center}

The amount of compression that is achieved depends greatly on the
particulars of the data contained in the table. However, in nearly every
case, transposing the rows and columns in the table (using GZIP\_1)
significantly improves the compression ratio, and shuffling the bytes in
the numeric columns (using GZIP\_2) provides even further improvement. 
The only real exception is file 11, where shuffling the bytes results in
worse compression.  This particular table is unusual, however, because the
data values in most of the columns are  nearly the same  in every row.
Gzip is apparently better able to compress  these repetitive  values if
the individual bytes are not shuffled.

The last column in the table  gives the ratio of the amount of disk space
that is saved when compressing the table using the GZIP\_2 method versus
just using the gzip utility program to compress the whole table.  On
average, GZIP\_2 saves 1.5 times more disk space.

Using this current prototype implementation, it takes about 50\%  more
CPU time to compress the table using the GZIP\_2 method than when using the
gzip utility.  It is expected, however, that further optimization of the 
code will eliminate most of this time difference.

Finally, it should be noted that these tests were performed on large FITS
tables where the size of the required FITS primary array header, plus the
size of the table header, plus the size of any fill bytes at the end of  the
last FITS 2880-byte block of the table, is insignificant compared to the
size of the data table itself.  Because this tiled-table compression method
does not compress the headers, and because the minimum size of the compressed
table is 1 FITS block (2880 bytes), this compression method will be less
effective for small tables or for table where the headers account for a
significant fraction of the size of the table.

We anticipate that further prototyping work on this table
compression convention will be performed in the near future.  In
particular, the performance of other lossless compression algorithms 
beside gzip should to be explored.  It may also be desirable to consider
extending this convention to support compression of ASCII FITS tables,
and possibly add an option for compressing the headers of images or tables
that contain a relatively large number of keywords.

\smallskip
\centerline{\bf References}

{\parskip=0pt \frenchspacing \advance\leftskip by\parindent \parindent=-\parindent

HDF 2000, ``Performance Evaluation Report: gzip, bzip2 compression with and without
shuffling,''  \verb+http://www.hdfgroup.org/HDF5/doc_resource/H5Shuffle_Perf.pdf+
(not to be confused with \verb+http://www.youtube.com/watch?v=YjfHRRrFUPk+)

Rice, R. F., Yeh, P.-S., and Miller, W. H.
1993,  in Proc. of the 9th AIAA Computing in Aerospace Conf.,
AIAA-93-4541-CP, American Institute of Aeronautics and Astronautics

White, R. L., Greenfield, P., Pence, W.,  Tody, D.,
and Seaman, R. 2009, ``Tiled Image Compression Convention'', 
\verb+http://fits.gsfc.nasa.gov/registry/tilecompression.html+

}
\end{document}